\documentclass[10pt]{article}

\def \bsr {(b,\sigma,r)}

\def \D {\hbox{d}}
\def \Log {\mathop{\rm Log}\nolimits}

\begin{document}

\title{Letter to the Editor\\
\textbf{Another integrable case in the Lorenz model}}

\author
{
Tat-Leung Yee\footnote{
Permanent address:
 Department of Mathematics,
 The Hong-Kong university of science and technology,
 Clear Water Bay, Kowloon,
 Hong Kong.
\hfill S2004/003.
}
\ and
Robert Conte
\\
Service de physique de l'\'etat condens\'e
(URA 2464)
\\~~CEA--Saclay, F--91191 Gif-sur-Yvette Cedex, France
\\[10pt]
E-mail: TonYee@ust.hk and Conte@drecam.saclay.cea.fr
}

\maketitle

\hfill 

{\vglue -10.0 truemm}
{\vskip -10.0 truemm}

\begin{abstract}
A scaling invariance in the Lorenz model allows one to consider
the usually discarded case $\sigma=0$. 
We integrate it with the third Painlev\'e function.
\end{abstract}

\noindent \textit{Keywords}:
Lorenz model,
first integral,
third Painlev\'e equation.

\noindent \textit{PACS 1995}~:
 02.30.-f,  
 05.45.+b,  
 47.27.-i,  

\baselineskip=12truept


\section{Introduction}

The Lorenz model \cite{Lorenz}
\begin{eqnarray}
\frac{\D x}{\D t} = \sigma (y-x),\
\frac{\D y}{\D t} = r x - y - x z,\
\frac{\D z}{\D t} = x y - b z,
\label{eqLorenz}
\end{eqnarray}
in which $\bsr$ are real constants,
is a prototype of chaotic behaviour  \cite{GH}.
In particular, it fails the Painlev\'e test unless the parameters 
obey the constraints \cite{Segur}
\begin{eqnarray}
Q_2 & \equiv & (b-2 \sigma) (b + 3 \sigma -1)=0,
\label{eqQ2}
\\
\forall x_2:\ Q_4 & \equiv &
- 4 i (b - \sigma -1) (b - 6 \sigma + 2) x_2
-(4/3) (b - 3 \sigma +5) b \sigma r
\nonumber
\\
& &
+(-4 + 10 b + 30 b^2 - 20 b^3 -16 b^4)/27
\nonumber
\\
& &
+(-38 b - 56 b^2 -(28/3) b^3 + 88 \sigma + 86 b^2 \sigma) \sigma /3
\nonumber
\\
& &
- 32   \sigma/9
+ 70 b \sigma^2
- 64   \sigma^3
- 58 b \sigma^3
+ 36   \sigma^4
=0.
\label{eqQ4}
\end{eqnarray}
This system (\ref{eqQ2})--(\ref{eqQ4}) depends on $r$ only through the
product $b \sigma r$,
as a consequence of an obvious scaling invariance in the model,
and it admits four solutions,
\begin{eqnarray}
& &
(b,\sigma,b \sigma r)=
(1,1/2,0  ),
(2,1  ,2/9),
(1,1/3,0),
(1,0  ,0).
\end{eqnarray}
In the first three cases, 
i.e.~when the system (\ref{eqLorenz}) is nonlinear, 
which excludes $\sigma=0$,
the system can be explicitly integrated \cite{Segur},
and the general solution $(x,y,z)$ is a singlevalued function of time 
expressed with, respectively,
an elliptic function, the second and the third Painlev\'e functions.

In this letter,
we consider the fourth case
\begin{eqnarray}
& &
\bsr=
(1,0  ,r  ).
\end{eqnarray}
The apparently linear nature of the dynamical system can be removed
by eliminating $y$ and $z$ and considering
the third order differential equation for $x(t)$ \cite{SenTabor},
\begin{eqnarray}
& &
y = x + x' / \sigma,\
z = r-1-[(\sigma+1) x' + x''] / (\sigma x),
\\
& &
x x''' - x' x'' + x^3 x'
+ \sigma x^4 + (b + \sigma + 1) x x'' + (\sigma + 1) (b x x' - x'^2)
\nonumber
\\
& &
+ b (1-r) \sigma x^2 = 0,
\label{eqLorenzODEx}
\end{eqnarray}
which also depends on $r$ only through the product $b \sigma r$,
and thus implements the above mentioned scaling invariance.
The necessary conditions for (\ref{eqLorenzODEx}) to pass
the Painlev\'e test are the same ($Q_2=0,Q_4=0$) 
as for the dynamical system (\ref{eqLorenz}),
the restriction $\sigma\not=0$ being now removed.

\section{Integration for $b=1,\sigma=0$}

Because of the scaling invariance, the following first integral \cite{Segur}
of the dynamical system (\ref{eqLorenz}),
\begin{eqnarray}
& &
\bsr=(1,\sigma,0):\
K_3=(y^2 + z^2) e^{2 t},
\end{eqnarray}
is also a first integral of the third order equation 
for $(b,\sigma,b \sigma r)=(1,\sigma,0)$,
which includes the particular case of interest to us
$(b,\sigma,b \sigma r)=(1,0,0)$,
\begin{eqnarray}
& &
(b,\sigma,b \sigma r)=(1,0,0):\
K^2=\lim_{\sigma \to 0} \sigma^2 K_3 =
\left[\left(\frac{x''+x'}{x}\right)^2 +{x'}^2\right] e^{2t}.
\end{eqnarray}
For $K=0$, the general solution is
\begin{eqnarray}
& &
x=i k \tanh \frac{k}{2} (t-t_0) -i,\ i^2=-1,\ (k,t_0) \hbox{ arbitrary}.
\end{eqnarray}
For $K\not=0$,
after taking the usual parametric representation
\begin{eqnarray}
& &
\frac{x''+x'}{x}= K e^{-t} \cos \lambda,\
x'              = K e^{-t} \sin \lambda,
\end{eqnarray}
the second order ODE for $\lambda(t)$ is found to be
\begin{eqnarray}
& &
\lambda'' - K e^{-t} \sin \lambda=0,
\label{eqlambda}
\end{eqnarray}
with the link
\begin{eqnarray}
& &
x(t)=\lambda'(t).
\end{eqnarray}
In the variable $\cos \lambda$,
the differential equation (\ref{eqlambda}) becomes algebraic 
and belongs to an already integrated class \cite{GambierThese}.
The overall result is the general solution
\begin{eqnarray}
& &
x=i+2 i \frac{\D}{\D t} \Log w(\xi(t)),\ i^2=-1,\
\xi=a e^{-t},\
\end{eqnarray}
in which $w(\xi)$ is the particular third Painlev\'e function defined by
\begin{eqnarray}
& &
\frac{\D^2 w}{\D \xi^2}=\frac{1}{w} \left(\frac{\D w}{\D \xi}\right)^2
 - \frac{\D w}{\xi \D \xi}
+ \frac{\alpha w^2 + \gamma w^3}{4 \xi^2}
+\frac{\beta}{4 \xi} + \frac{\delta}{4 w},
\\
& &
\alpha=0,\
\beta=0,\
\gamma \delta= - (K/a)^2.
\end{eqnarray}

\section{Conclusion}

Out of the two cases selected by the condition $Q_2=0$,
one admits a first integral \cite{Segur},
\begin{eqnarray}
& &
b=2 \sigma:\ K_1=(x^2-2 \sigma z)e^{2 \sigma t},
\end{eqnarray}
but, in the second case $b=1-3 \sigma$,
the first integral whose existence has been conjectured \cite{TW} 
is not yet known.
The present result, which belongs to this unsettled case $b=1-3\sigma$,
should help to solve this open question.

\section*{Acknowledgments}

T.-L.~Yee thanks the Croucher Foundation for a postdoc grant at CEA.
R.~Conte acknowledges the support of the 
France-Hong Kong PROCORE grant 04807SF,
which allowed starting this work.


\vfill \eject
\end{document}